# Design of a hybrid plasmonic electro-optical modulator based on n-doped silicon and barium titanate

Purya Es'haghi[a]* and Ali Barkhordari[b]

[a]*Department of Photonics, Graduate University of Advanced Technology, kerman, iran;*
[b]*Department of plasma engineering, Graduate University of Advanced Technology, kerman, iran*

[a]*E-mail: puryae@gmail.com

[a]*E-mail: p.eshaghi@student.kgut.ac.ir

[a]*ORCID ID: https://orcid.org/0000-0002-2203-9497

[a]*LinkedIn: https://www.linkedin.com/in/purya-es-haghi-a96154a7/

[b]E-mail: alibarkhordari20@yahoo.com

Purya E'shaghi studied the solid state physics in Bsc at 2013 and graduated photonics in Msc at 2016. He is interested in researching on plasmonic modulators and geraphene based plasmonic sensors and plasmonic waveguides.

Ali Barkhordari educated the Optics and laser engineering in Bsc at 2013 and graduated the plasma engineering in Msc at 2016. He is interested in investigating on plasma spectroscopy, plasma simulation, laser spectroscopy, optics and laser instruments, plasma tools, laser-plasma interaction, plasmonic modulators.

# Design of a hybrid plasmonic electro-optical modulator based on n-doped silicon and barium titanate

**Abstract:**

In this paper, a numerical solution for a hybrid plasmonic modulator is presented with a six-layer structure consisting of an air superstrate, a gold layer, a barium titanate layer, a n type silicon layer, a gold layer and an $Al_2O_3$ nanolattice substrate. Regarding the suggested structure, the parameters related to the phase and the absorption modulation are investigated at different thicknesses. Here, the Pockels effect and the free carrier dispersion effect are considered simultaneously. The dispersion equation of this structure is analytically obtained and numerically solved by the Nelder-Mead method. The minimum π shift length is predicted to be equal to 6.91μm and the maximum calculated figure of merit is 10.74. Furthermore, according to our results, it is understood that this modulator has a high ability to be utilized in optical communication systems. Also, it could be integrated to the microelectronic systems and it is compatible with CMOS technology.

## 1. Introduction

According to Moore's law, the number of transistors in a fixed area of dense integrated circuit doubles almost every two years and consequently electronic devices are going to be more compact. Because of the diffraction limit, optical waveguides do not have compaction capability below the half of propagation wavelength and hence they can't be integrated with electronic devices. This is a defect in optical waveguides. Recently, scientists have turned to plasmonic waveguides in order to overcome the diffraction limit [1-3].

Meanwhile, it is important to use the plasmonic waveguide-based modulators and switches. Because these modulators and switches have footprints on the order of several μm², they have the capability to cointegrate with electronic systems [4,5]. These types of modulators and switches have many advantages, such as high bandwidth [6], low power consumption [7], and ultra-compact size [8]. Also these mainly employ the thermal optics effect [9-11], the free carrier dispersion effect [12-14], the Pockels effect [15-17], the phase transition effect [18-20] and the electrochemical metallization effect [21,22]. Owing to the usage of different types of waveguides based on surface plasmon polaritons, these devices have a strong interaction between propagating optical signals and active materials [23-28], which is necessary to produce optical switches well.

However, plasmonic modulators and switches suffer some disadvantages and deficiencies, including low propagation length and high propagation loss. To overcome these obstacles, scientists have invented hybrid plasmonic waveguides. Combination of the optical and the plasmonic properties in these types of waveguides, cause to achieve some special benefits, such as high optical limitations, low propagation losses and its updatable sufficiency with photonic devices [29]. Among the waveguide-based modulators, silicon-based is the most important one. Forasmuch as it is compatible with CMOS technology, it's able to integrate with microelectronics devices.

Here, a hybrid plasmonic modulator is offered, which has an Air/Au/BaTiO$_3$/n-Si/Au/Al$_2$O$_3$ nanolattice with near unity refractive index configuration. This structure is studied as an intensity modulator and a phase modulator at 1.55μm wavelength. In order to analyze this structure's properties, the dispersion equation is obtained and solved for 0 and 10V applied voltages. Therefore, the Poisson equation is solved to apply an electric field to the voltage-dependent variation of the refractive index. It should be noted that the free carrier dispersion effect and the Pockels effect are simultaneously considered in this device.

The different sections of this paper are organized as follows: In section 2, the simulation theory is discussed. The theory includes how to govern the dispersion and Poisson equations, as well as numerical methods for solving them. Moreover, the investigated parameters in the phase and intensity modulation format are presented in this section. In section 3, results of the simulation are shown and the performance of the modulator is analyzed in phase and intensity modulation. In section 4, the final conclusion of this paper is expressed.

## 2. Theory

### *2.1. Modulator Structure*

The studied modulator structure is a slab waveguide. This waveguide consists of two layers of n-Si and $BaTiO_3$ which are embedded between two layers of Au. The substrate of this structure is a nanolattice of $Al_2O_3$ with near unity refractive index. The above region of the structure is filled with the air. The $BaTiO_3$ is a birefringent crystal that matches its principal-axis system to the axes of the considered coordinate system. Figure 1 shows the structure of the investigated modulator.

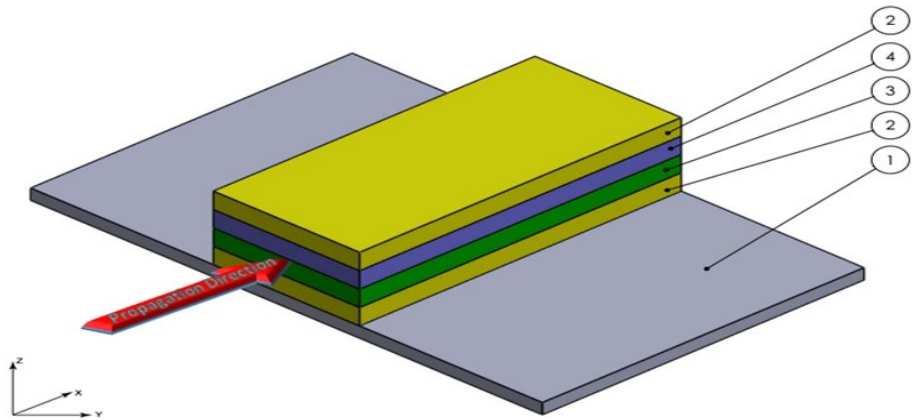

Figure 1. Schematic structure of designed modulator; (1) nanolattice of $Al_2O_3$ with near unity refractive index substrate (2) gold layer (3) n-Si layer (4) $BaTiO_3$ layer.

The refractive indices of materials, which form the modulator, can be defined as:

for substrate [30]:

$$n_{Al_2O_3} = 1.025 \tag{1}$$

for two layers of Au [31]:

$$n_{Au}^2 = 1.53 - \frac{1}{145^2\left(\frac{1}{\lambda^2}+\frac{i}{17000\lambda}\right)} + \frac{0.94}{468}\left[\frac{e^{-i\pi/4}}{\left(\frac{1}{468}-\frac{1}{\lambda}-\frac{i}{2300}\right)} + \frac{e^{-i\pi/4}}{\left(\frac{1}{468}+\frac{1}{\lambda}+\frac{i}{2300}\right)}\right] + \frac{1.36}{331}\left[\frac{e^{-i\pi/4}}{\left(\frac{1}{331}-\frac{1}{\lambda}-\frac{i}{940}\right)} + \frac{e^{-i\pi/4}}{\left(\frac{1}{331}+\frac{1}{\lambda}+\frac{i}{940}\right)}\right] \tag{2}$$

and for the n-Si [32]:

$$\varepsilon_{n-Si} = \varepsilon_\infty - \frac{\omega_p^2}{[\omega^2(1+\frac{1}{\omega\tau})]}, \quad \omega_p^2 = \frac{Ne^2}{\varepsilon_0 m^*}, \quad \tau = \frac{m^*\mu}{e} \tag{3}$$

where $\mu = 50 \text{ cm}^2/\text{Vs}$, $m_e = 9.1 \times 10^{-31} \text{kg}$, $m^* = 0.272 m_e$, $\varepsilon_\infty = 11.7$, $e = 1.602 \times 10^{-19}\text{C}$ and $\varepsilon_0 = 8.85 \times 10^{-12} \text{ C}^2/\text{Nm}^2$ are considered. Also, the ordinary and extraordinary refractive indexes of the BaTiO3 can be written as [33]:

$$n_o^2 = 3.05840 + \frac{2.27326\lambda^2}{\lambda^2 - 0.07409} - 0.02428\lambda^2 \tag{4}$$

$$n_e^2 = 3.02479 + \frac{2.14062\lambda^2}{\lambda^2 - 0.067007} - 0.02169\lambda^2 \tag{5}$$

In equations (1) to (5), n, λ, $\varepsilon_\infty$, $\omega_p$, τ, μ, $m_e$, $m^*$, $\varepsilon_0$, e and N are the refractive index, the wavelength, the background permittivity, the plasma frequency, the relaxation time, the electron mobility, the electron mass, the electron effective mass, the vacuum permittivity, the electron charge and the free carrier density, respectively. The free carrier density is approximately equal to the doping concentration in the heavily doped n-Si. In this case the doping concentration is equal to $10^{21} cm^{-3}$ [34].

Surface Plasmon-Polaritons (SPPs) are propagated between two gold layers. By applying a voltage between two layers of gold, a static electric field is created inside the dielectric layer. This field changes the refractive index of BaTiO3, it also causes a shift in the real and imaginary parts of the n-Si refractive index. These variations also lead to alter the intensity of surface plasmon polaritons. To investigate these evolutions, it's necessary to obtain the dispersion equation.

*2.2. Dispersion equation*

In order to determine the dispersion equation for the structure in the on and off states, the TM mode solutions are used in the wave equation. By assuming that the time dependence part of electromagnetic field is $e^{-i\omega t}$, the wave equation for the magnetic field component can be described by [35]:

$$\vec{\nabla} \times (\overleftrightarrow{\varepsilon}^{-1} \vec{\nabla} \times \vec{H}) - \omega^2 \mu_0 \vec{H} = 0 \tag{6}$$

As it is required to stimulate the TM modes for SPP modes, thus the wave equation is indicated as [36]:

$$\frac{\partial^2 H_y}{\partial z^2} + \left(k_0^2 \varepsilon_x - \beta^2 \frac{\varepsilon_x}{\varepsilon_z}\right) H_y = 0 \tag{7}$$

The components of the electric field for the TM mode are achieved as:

$$E_x = \frac{-i}{\omega \varepsilon_0 \varepsilon_x} \frac{\partial H_y}{\partial z} \tag{8}$$

$$E_z = \frac{-\beta}{\omega \varepsilon_0 \varepsilon_z} H_y \tag{9}$$

The n-Si layer is divided into 10 parts for the more accurate study of electrostatic and electrodynamic field effects. By assuming that the time and the x-coordinate dependent parts are in $e^{-i(\omega t-\beta x)}$ form, the solutions which satisfy equations (7), (8) and (9), are as:

$$H_y = \begin{cases} A_1 e^{ik_1(z-z_1)} & -\infty < z < z_1 \\ A_2 e^{ik_2(z-z_2)} + A_3 e^{-ik_2(z-z_2)} & z_1 < z < z_2 \\ A_4 e^{ik_3(z-z_3)} + A_5 e^{-ik_3(z-z_3)} & z_2 < z < z_3 \\ A_6 e^{ik_4(z-z_4)} + A_7 e^{-ik_4(z-z_4)} & z_3 < z < z_4 \\ \vdots & \vdots \\ A_{22} e^{ik_{12}(z-z_{12})} + A_{23} e^{-ik_{12}(z-z_{12})} & z_{12} < z < z_{13} \\ A_{24} e^{ik_{13}(z-z_{13})} + A_{25} e^{-ik_{13}(z-z_{13})} & z_{13} < z < z_{14} \\ A_{26} e^{ik_{14}(z-z_{14})} & z_{14} < z < +\infty \end{cases} \quad (10)$$

where $z_1 = 0$, $z_2 = h_{Au}$, $z_3 = h_{Au} + \frac{h_{n-Si}}{10}$, $z_4 = h_{Au} + \frac{2h_{n-Si}}{10}$, ... , $z_{12} = h_{Au} + \frac{10h_{n-Si}}{10}$, $z_{13} = h_{Au} + h_{n-Si} + h_{BaTiO_3}$, , $z_{14} = h_{Au} + h_{n-Si} + h_{BaTiO_3}$. By inserting equation (10) into equations (8) and (9), the components of the electric field of the TM mode are also obtained. In relation (10), k is the transverse propagation constant, which for all layers except the BaTiO3 layer, is equal to:

$$k_m = \sqrt{\varepsilon_m k_0^2 - \beta^2} \quad (11)$$

due to $\varepsilon_x = \varepsilon_z$, and for the BaTiO3 layer is equal to [36]:

$$k_{13} = \sqrt{\varepsilon_x k_0^2 - \beta^2 \frac{\varepsilon_x}{\varepsilon_z}} \quad (12)$$

where β is the propagation constant of SPP mode and $\varepsilon_x$ and $\varepsilon_z$ are the permittivities of the anisotropic layer in the x and z directions respectively also $k_0$ is the wavenumber in the free space. By applying the boundary conditions to $H_y$ and $E_z$, the following equation can be obtained as:

$$M_{28\times28}(\lambda, \beta) \cdot A_{28\times1} = 0 \tag{13}$$

$\lambda$ is the vacuum wavelength and $A_{28\times1}$ is a column matrix that describes the constants in front of the exponential functions in relation (10). This matrix can be defined as follows:

$$A = \begin{pmatrix} A_1 \\ A_2 \\ A_3 \\ . \\ . \\ . \\ A_{28} \end{pmatrix} \tag{14}$$

In order to distinguish the dispersion equation, the determinant of $M_{28\times28}$ should be calculated as [37]:

$$\det(M_{28\times28}(\lambda, \beta)) = |M_{28\times28}(\lambda, \beta)| = 0 \tag{15}$$

The propagation constant is acquired by solving the dispersion equation per wavelength. The resultant propagation constant will be a complex number.

*2.3. Solving method*

Here, the Nelder-Mead method is used to solve the dispersion equation [38]. The standard error to stop the numerical calculation is $\epsilon = 2.2204 \times 10^{-16}$. The absorption coefficient is calculated from the imaginary part of the propagation constant as:

$$\alpha = 2\text{Im}[\beta] \tag{16}$$

and then the effective propagation length is expressed as:

$$L_e = \frac{1}{\alpha} \tag{17}$$

The extinction ratio is characterized as:

$$E_R = (\log_{10} e)\Delta\alpha L \,.\, \Delta\alpha = |\alpha_{off} - \alpha_{on}| \tag{18}$$

Using the equation (18), the 1dB on-off length can be defined where the extinction ratio is equal to 1dB in an arbitrary chosen DC voltage that can be written as follows:

$$L_{1dB} = \frac{1}{(\log_{10} e)\Delta\alpha} \tag{19}$$

At last, it is worthy to illustrate a parameter for comparing the efficiency of the absorption plasmonic modulator based on the above-defined quantities, called the figure of merit (FoM) [37,39]:

$$FoM = \frac{L_e}{L_{1dB}} = (\log_{10} e)\frac{|\alpha_{off} - \alpha_{on}|}{\alpha_{state}} \tag{20}$$

Furthermore, it is better to demonstrate the π shift length for phase modulators. The length that waveguide needs to generate the π radians phase difference between the on and off modes is called π shift length, which is defined as:

$$L_\pi = \frac{\pi}{|\beta_{off} - \beta_{on}|} \tag{21}$$

where, in relation (18) to (21), $\Delta\alpha = |\alpha_{off} - \alpha_{on}|$ and $\alpha_{state}$, $\alpha_{off}$, $\alpha_{on}$, $\beta_{on}$, $\beta_{off}$ and L are the difference of the absorption coefficients, the residual absorption coefficient, the absorption coefficient of the off state, the absorption coefficient of the on state, the propagation constant in the on state, the propagation constant in the off state and the system length, respectively [40].

## 2.4. Applied static electric field

In general, for applying the static electric field in an anisotropic media, the nonlinear Poisson equation should be considered [41]:

$$\vec{\nabla} \cdot \overleftrightarrow{\varepsilon_s} \vec{\nabla} \varphi = -\frac{\rho}{\varepsilon_0} \quad (22)$$

As the charge density is zero ($\rho = 0$) in the BaTiO$_3$ layer, the equation (22) is reduced to Laplace equation. Nonlinear Poisson equation of this case is solved in one dimension. boundary conditions to acquire the electric potential ($\varphi$) at interfaces should be indicated:

$$\frac{U}{d} = \varepsilon_{s,z} \frac{d\varphi}{dz}\bigg|_{z \to +(h_{Au}+h_{n-Si})} = \varepsilon_{s,n-Si} \frac{d\varphi}{dz}\bigg|_{z \to -(h_{Au}+h_{n-Si})} \quad (23)$$

$$\varphi(z) = 0 \quad \text{for} \quad z \to +h_{Au} \quad (24)$$

$$\varphi(z) = U \quad \text{for} \quad z = h_{Au} + h_{n-Si} + h_{BaTiO_3} \quad (25)$$

In the n-Si layer, via exploiting the Thomas- Fermi screening theory, the Poisson equation can be rewritten as:

$$\nabla^2 \varphi = \frac{e(N_i(z)-N_0)}{\varepsilon_0 \varepsilon_{s,n-Si}} \quad (26)$$

In equation (26),

$$N_i(z) = \frac{1}{3\pi^2} \left(\frac{8\pi^2 m^*}{h^2}\right)(E_f + e\varphi(z)) \quad (27)$$

$$\varepsilon_{s,n-Si} = 11.688 + \frac{1.635 \times 10^{-19} N_D}{1+1.172 \times 10^{-21} N_D} \quad (28)$$

$$\varepsilon_{s,z} = 135 \quad (29)$$

and in relation (27), Fermi energy is defined as:

$$E_f = \left(\frac{h^2}{8\pi^2}\right)[3\pi^2 N_0]^{2/3} \tag{30}$$

From relation (22) to (30), $N_0$ is the bulk free carrier density of the n-Si, $N_D$ is the concentration of donors in the n-Si, $\varepsilon_{s,n-Si}$ is the relative static permittivity of the n-Si, $m^*$ is the electron effective mass of the n-Si, $E_f$ is the Fermi energy of the n-Si, h is the Planck constant, $\varphi$ is the electrical potential, $N_i$ is the free carrier density of the n-Si, $\rho$ is the electrical charge density and $\varepsilon_{s,z}$ is the relative static permittivity of the BaTiO$_3$ in z-direction [37,42]. The finite difference method with the meshing number of 6000 is applied for solving the equation (26) [43].

To solve the Poisson equation in the n-Si layer, the boundary conditions (23) and (24) are used. By solving the Poisson equation in this layer, $N_i$ is obtained in 10 parts of the n-Si layer and thus by substituting $N_i$ in relation (3), the effects of applying voltage on the refractive index of the n-Si are investigated.

In the BaTiO$_3$ layer, the boundary conditions (23) and (25) is used to solve the Laplace equation (31) analytically. Accordingly, the obtained static electric field induces an alteration in the ordinary and extraordinary refractive index which can be characterized with:

$$\frac{d^2\varphi}{dz^2} = 0 \tag{31}$$

$$\begin{bmatrix} \Delta(\frac{1}{n^2})_1 \\ \Delta(\frac{1}{n^2})_2 \\ \Delta(\frac{1}{n^2})_3 \\ \Delta(\frac{1}{n^2})_4 \\ \Delta(\frac{1}{n^2})_5 \\ \Delta(\frac{1}{n^2})_6 \end{bmatrix} = \begin{bmatrix} r_{11} & r_{12} & r_{13} \\ r_{21} & r_{22} & r_{23} \\ r_{31} & r_{32} & r_{33} \\ r_{41} & r_{42} & r_{43} \\ r_{51} & r_{52} & r_{53} \\ r_{61} & r_{62} & r_{63} \end{bmatrix} \begin{bmatrix} E_x \\ E_y \\ E_z \end{bmatrix} = \overleftrightarrow{r} \cdot \vec{E} \qquad (32)$$

$$\overleftrightarrow{\eta} = [\frac{1}{n^2}] = \begin{bmatrix} \frac{1}{n_x^2} + \Delta(\frac{1}{n^2})_1 & \Delta(\frac{1}{n^2})_6 & \Delta(\frac{1}{n^2})_5 \\ \Delta(\frac{1}{n^2})_6 & \frac{1}{n_y^2} + \Delta(\frac{1}{n^2})_2 & \Delta(\frac{1}{n^2})_4 \\ \Delta(\frac{1}{n^2})_5 & \Delta(\frac{1}{n^2})_4 & \frac{1}{n_z^2} + \Delta(\frac{1}{n^2})_3 \end{bmatrix} \qquad (33)$$

$\overleftrightarrow{r}$ is the linear electro-optical coefficients matrix (Pockels), where described in the standard crystallographic coordinate system that the z-direction indicates the crystal optical axis. $r_{13} = r_{23} = 19.5[\frac{pm}{V}]$, $r_{33} = 97[\frac{pm}{V}]$, $r_{42} = r_{51} = 1640[\frac{pm}{V}]$ and the other coefficients are zeroes. $\overleftrightarrow{\eta}$ is the electro-optically perturbed impermeability tensor and $\Delta(\frac{1}{n^2})_m$ are the electro-optically induced variations of the impermeability tensor elements, also $n_x$, $n_y$, and $n_z$ are the refractive indices in x, y, and z-directions, respectively. The refractive indices resulting from the applied voltage are substituted into dispersion equation and it is solved again [44,45].

### 3. Results and Analysis

#### *3.1. Phase Modulation*

Both the real and the imaginary parts of the propagation constant alter by an applying voltage, consequently, it causes an enhancement of the free carrier density in the n doped silicon and an increase of the refractive index of the barium titanate. The variation in the real part of the propagation constant results in phase modulation.

Forasmuch as the most optical modulations lead to the intensity modulation ultimately, a Mach-Zender interferometer configuration is exploited to design the plasmonic phase modulator. In this way, to reach the off state, the π radians phase difference is created between the two arms and 2nπ (n is an integer number) radians phase difference is needed to apply between them to achieve the on state. To analyze the phase modulation, the considered parameters are the absorption coefficient, the propagation constant, the π shift length and the transmittance.

Figure 2 indicates that the absorption coefficient of the modulator is reduced by increasing the thickness of the gold layer. Also, by increasing the thickness of the $BaTiO_3$ layer, the absorption coefficient of this structure decreases. Moreover, it is obvious that the absorption coefficient raises at the off state. At the n-doped silicon layer thicknesses of 12nm and 13nm, by applying voltage, the absorption coefficient drastically alters with respect to the other thicknesses. At the 12nm thick n-doped silicon layer, the 30nm thick barium titanate layer and the 30nm thick gold layer, the absorption coefficient is at highest amount.

(a) 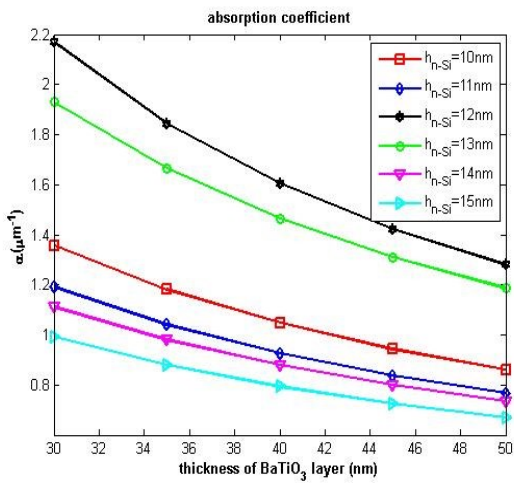

(b) 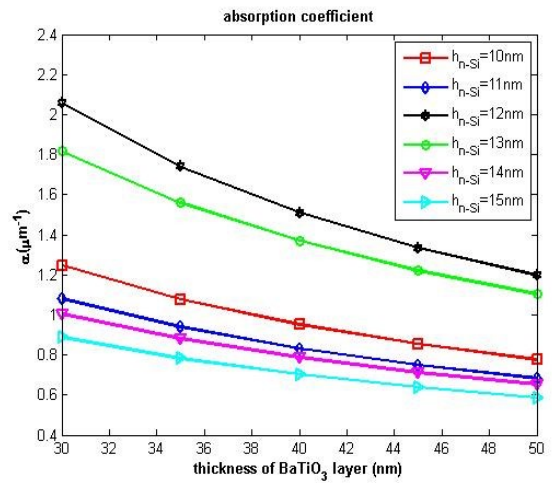

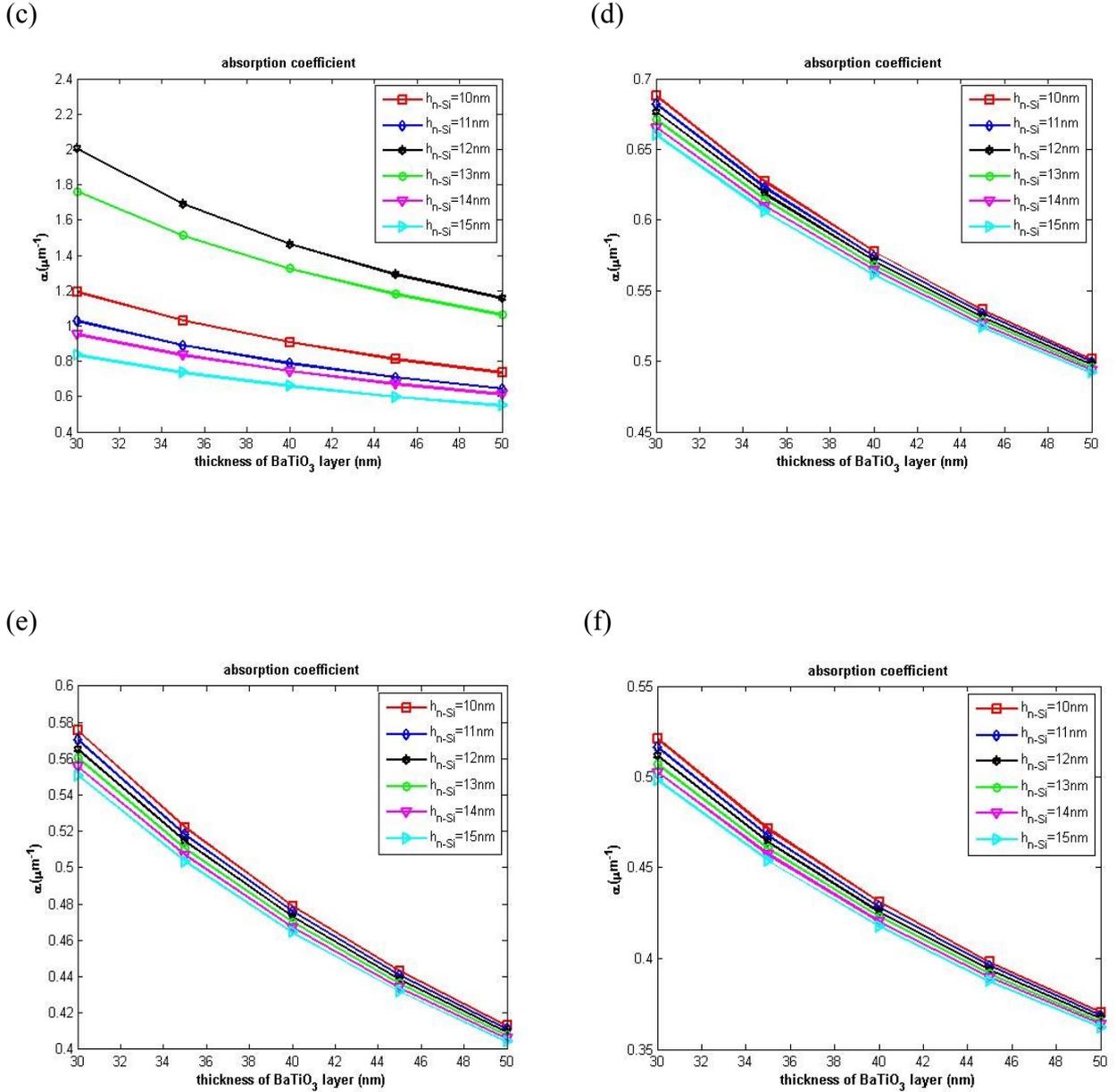

Figure 2. absorption coefficient in the off and on states in different thicknesses of the barium titanate and the n-Si, absorption coefficient in the off state in (a) $h_{Au}$=30 nm, (b) $h_{Au}$=40 nm, (c) $h_{Au}$=50 nm, absorption coefficient in the on state in (e) $h_{Au}$=30 nm, (f) $h_{Au}$=40 nm, (g) $h_{Au}$=50 nm.

Figure 3 depicts the alterations of the propagation constant as a function of the thickness of the n-doped silicon layer, the $BaTiO_3$ layer, and the gold layer. According to this figure, it can be found that by growing the thickness of the $BaTiO_3$, the propagation constant decreases in the on and off states. In the on state, the propagation constant has a low sensitivity to the thickness of the n-Si layer, but in the off state it changes noticeably. This parameter in the on and off states decreases with raising the thickness of the gold layer.

(a) (b)

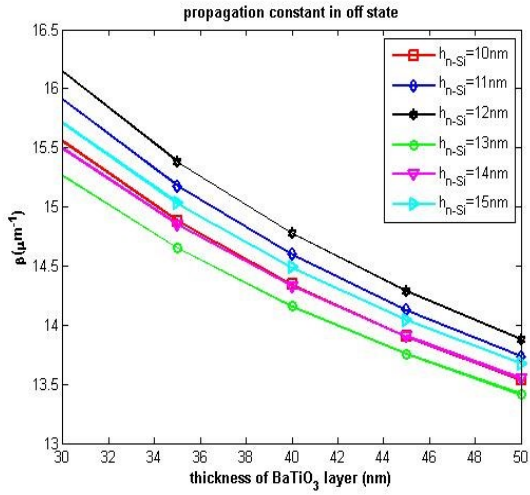 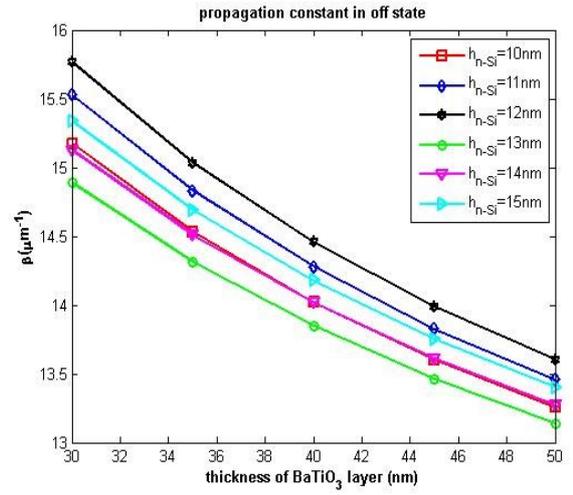

(c) (d)

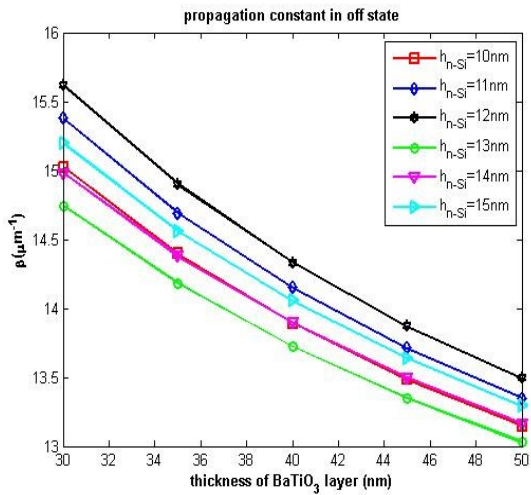 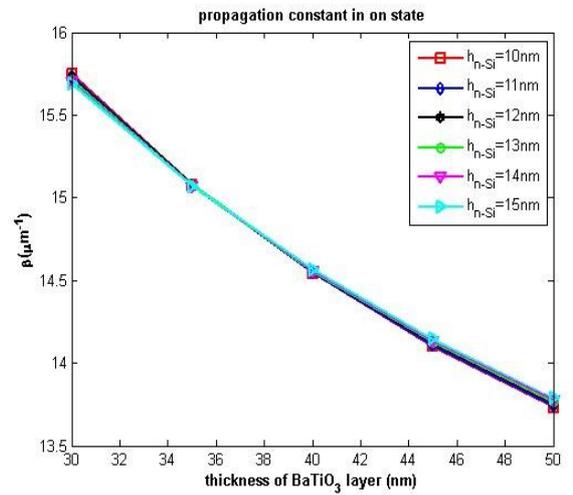

(e) 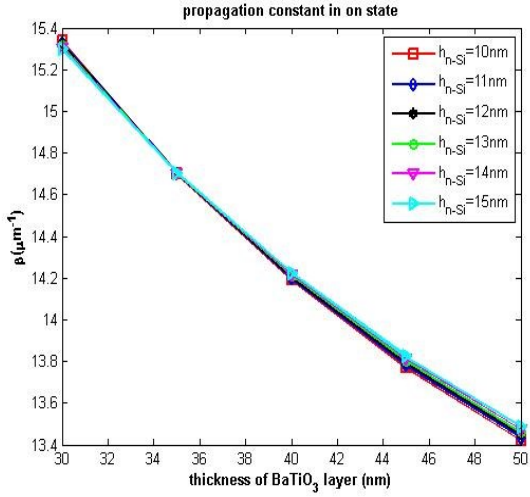  (f) 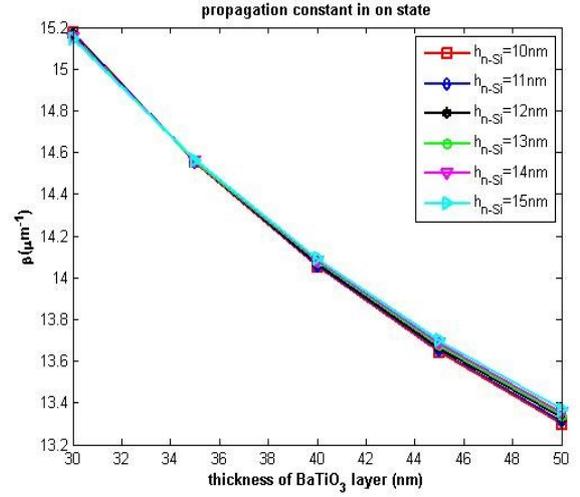

Figure 3. propagation constants in the off and on states in different thicknesses of the barium titanate and the n-Si, propagation constant in the off state in (a) $h_{Au}$=30 nm, (b) $h_{Au}$=40 nm, (c) $h_{Au}$=50 nm, propagation constant in the on state in (e) $h_{Au}$=30 nm, (f) $h_{Au}$=40 nm, (g) $h_{Au}$=50 nm.

The variation of the π shift length as a function of different thicknesses is shown in Figure 4. for the considered structure the lowest value for the π shift length is 6.91 at an Air(∞)/Au(50nm)/BaTiO$_3$(30nm)/n-Si(12nm)/Au(50nm)/Al$_2$O$_3$(∞) multilayer structure. Compared to the results of reference [39], it can be distinguished that smaller values for the π shift lengths can be obtained for our discussed structure, furthermore its value of calculated transmittance is -15.36 and -60.16 for the on and off states, respectively.

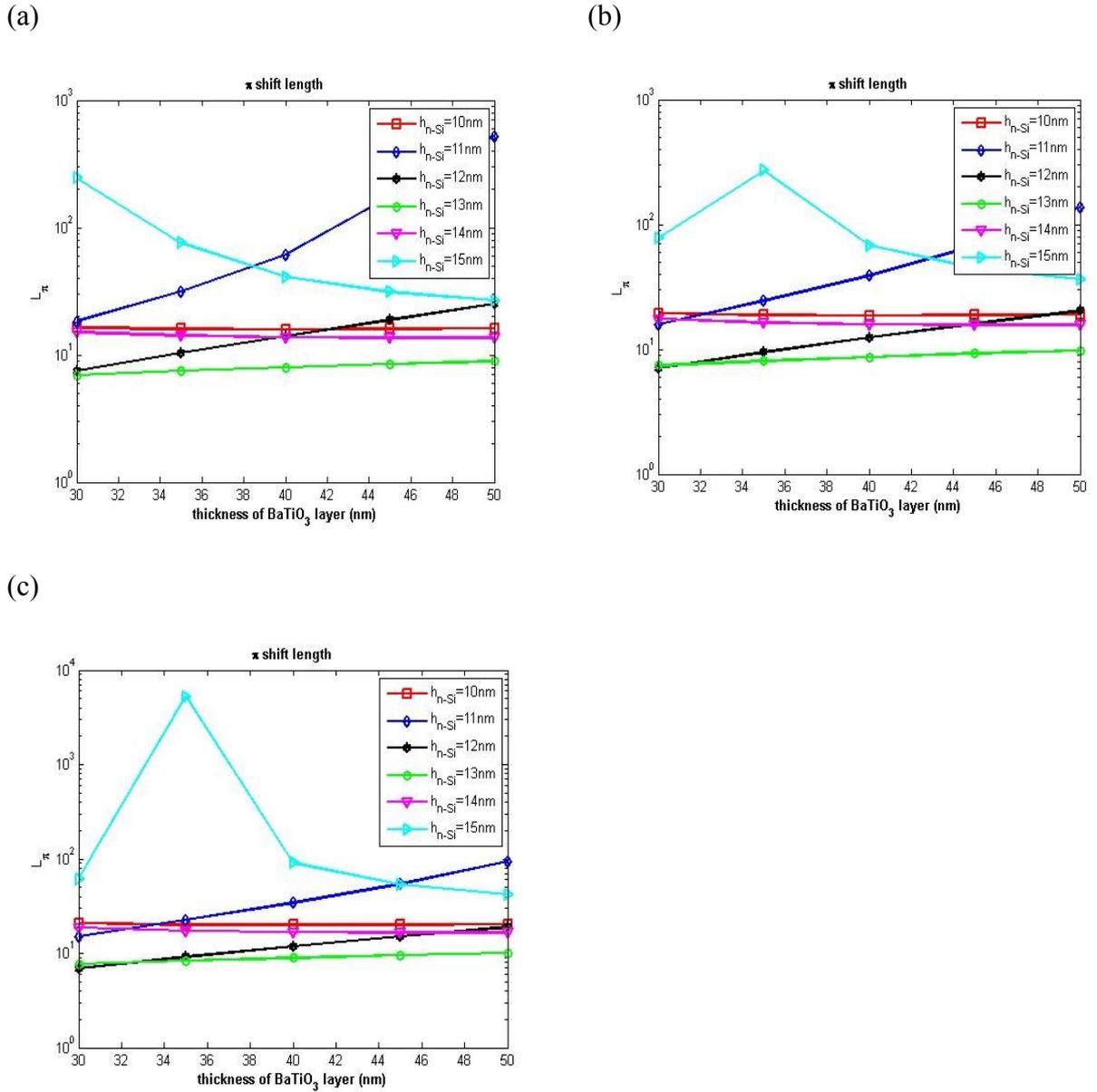

Figure 4. π shift length in (a) $h_{Au}$=30 nm, (b) $h_{Au}$=40 nm, (c) $h_{Au}$=50 nm.

The distribution of magnetic field, the electric field and the time-averaged of Poynting vector of the asymmetric plasmonic mode and the dispersion relation of this mode at the minimum value of the π shift length are plotted in Figure 5 and Figure 6 respectively. As regards to Figure 5, the distribution of the time-averaged Poynting vector increases with applying voltage, accordingly it leads to more sensitivity of the plasmonic mode with the variation of refractive indices.

(a) 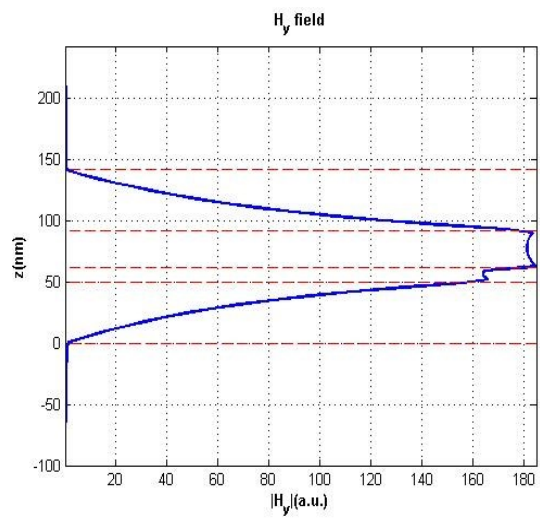

(b) 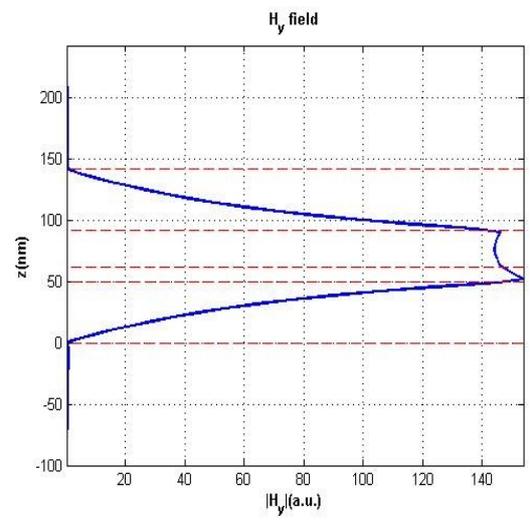

(c) 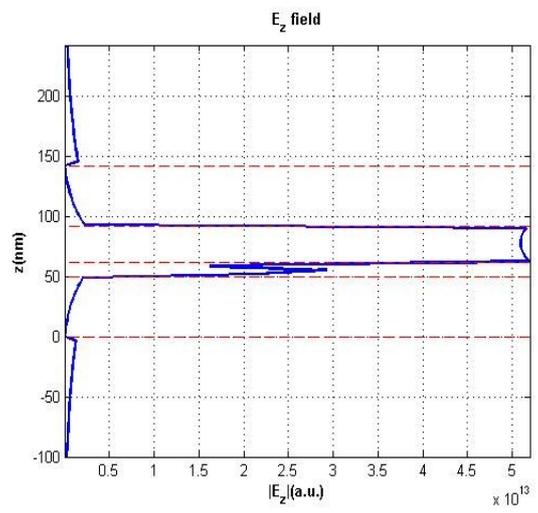

(d) 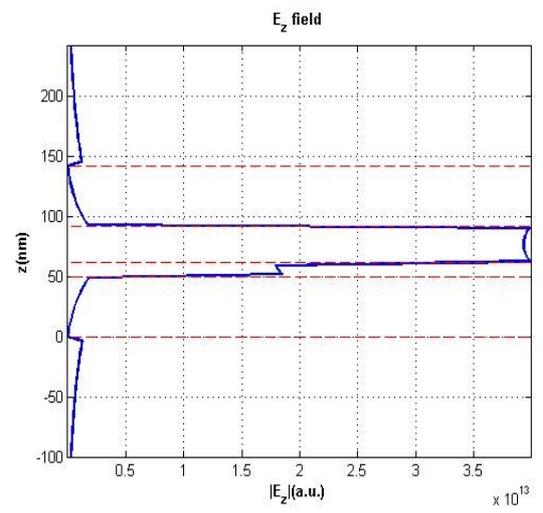

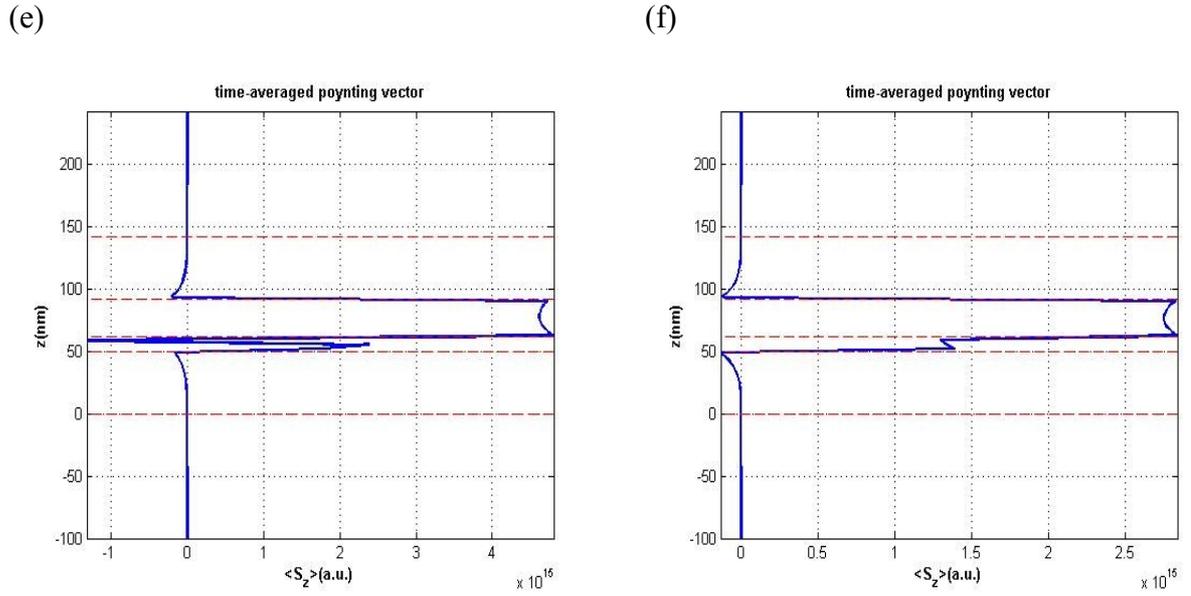

Figure 5. distribution of the magnetic field in (a) the off and (b) on states and distribution of electric field in (c) the off and (d) on states and the time-averaged Poynting vector in (e) the off and (f) on state of the Air(∞)/Au(50nm)/BaTiO$_3$(30nm)/n-Si(12nm)/Au(50nm)/Al$_2$O$_3$(∞) structure.

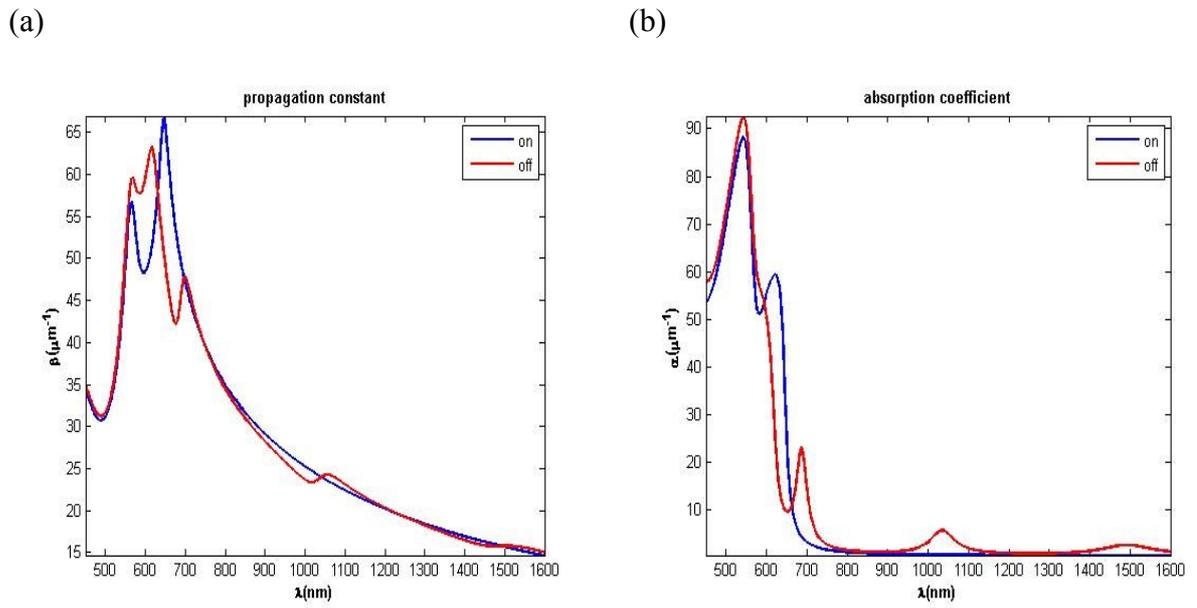

Figure 6. dispersion relation of the plasmonic mode in the Air(∞)/Au(50nm)/BaTiO$_3$(30nm)/n-Si(12nm)/Au(50nm)/Al$_2$O$_3$(∞) structure. (a) Propagation constant in the on and off states in different wavelengths, (b) absorption coefficient in the on and off state in different wavelengths.

## 3.2. Absorption Modulation

With regard to the variation of the imaginary part of the propagation constant of the proposed structure with applied voltage, it can also be utilized as an absorption modulator. The absorption coefficient of the structure switches the on state to the off state and it increases by applying voltage. In the absorption modulator, it is necessary to increase the effective propagation length and decrease the 1dB on-off length. thus the increase of figure of merit should be considered.

From the Figure 7, it can be seen that by increasing the thickness of the $BaTiO_3$ layer, the effective propagation length increases in the on and off states. Also, the maximum value of the effective propagation length of the suggested structure in the on state is 2.76μm, due to an Air(∞)/Au(50nm)/$BaTiO_3$(50nm)/n-Si(15nm)/Au(50nm)/$Al_2O_3$(∞) multilayer configuration. Furthermore, the effective propagation length will increase by growing the thickness of the gold layer.

(a) (b)

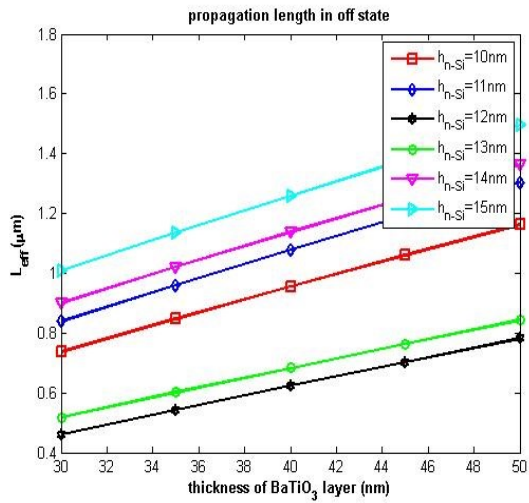 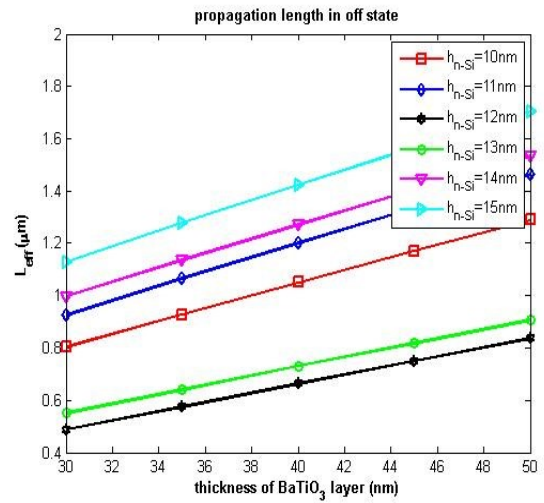

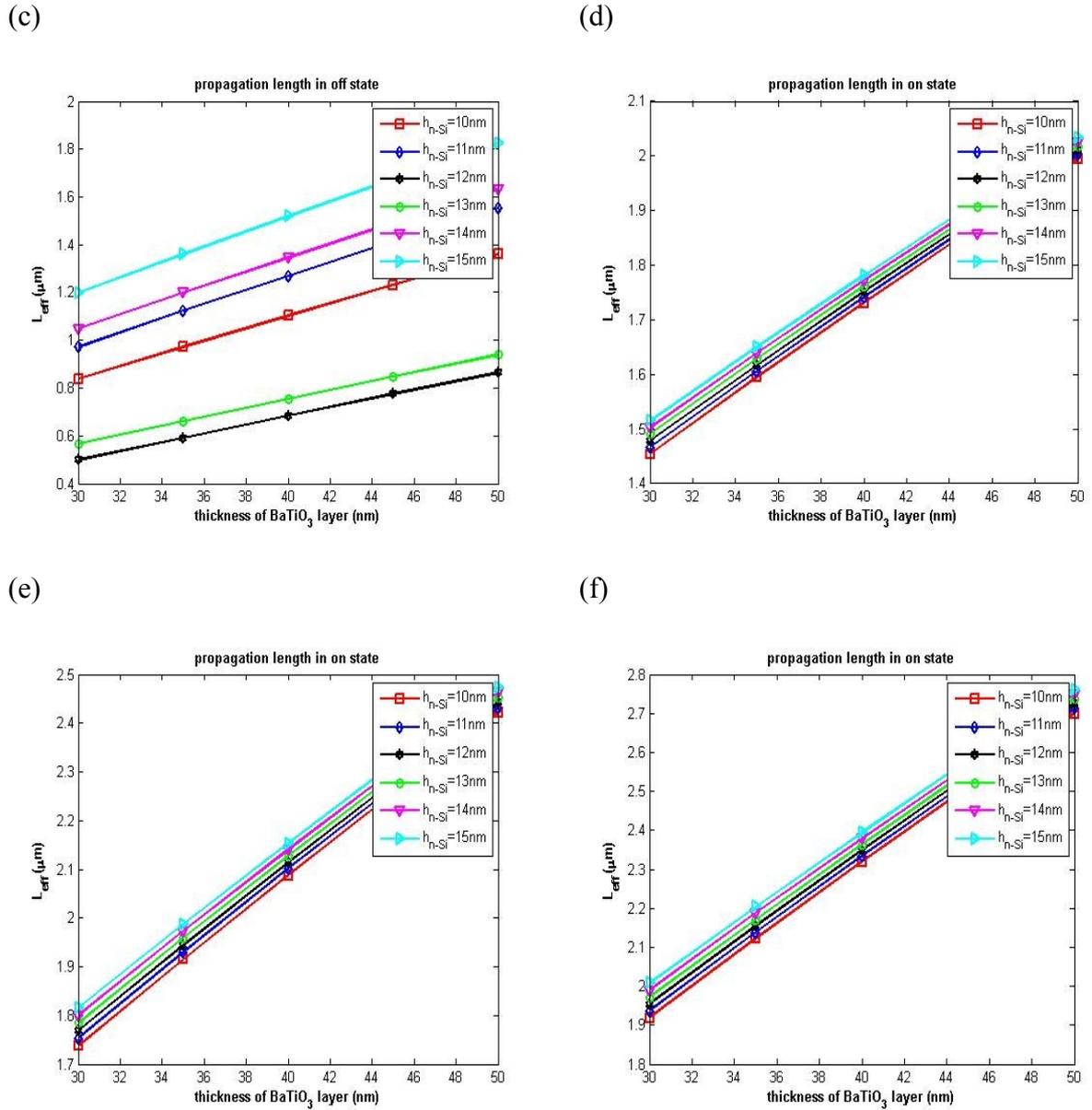

Figure 7. effective propagation length in different thicknesses. effective propagation length in the off state in (a) $h_{Au}$=30 nm, (b) $h_{Au}$=40 nm, (c) $h_{Au}$=50 nm, effective propagation length in the on state in (d) $h_{Au}$=30 nm, (e) $h_{Au}$=40 nm, (f) $h_{Au}$=50 nm.

Figure 8 shows the 1dB on-off length in various thicknesses. The minimum value of 1dB on-off length is 0.0407μm which occurred at an Air(∞)/Au(30nm)/BaTiO$_3$(50nm)/n-Si(15nm)/Au(30nm)/Al$_2$O$_3$(∞) multilayer structure. The 1dB on-off length decreases by increasing the BaTiO$_3$ thickness. Moreover, by growing the gold layer thickness, the 1dB on-off length has a negligible alteration.

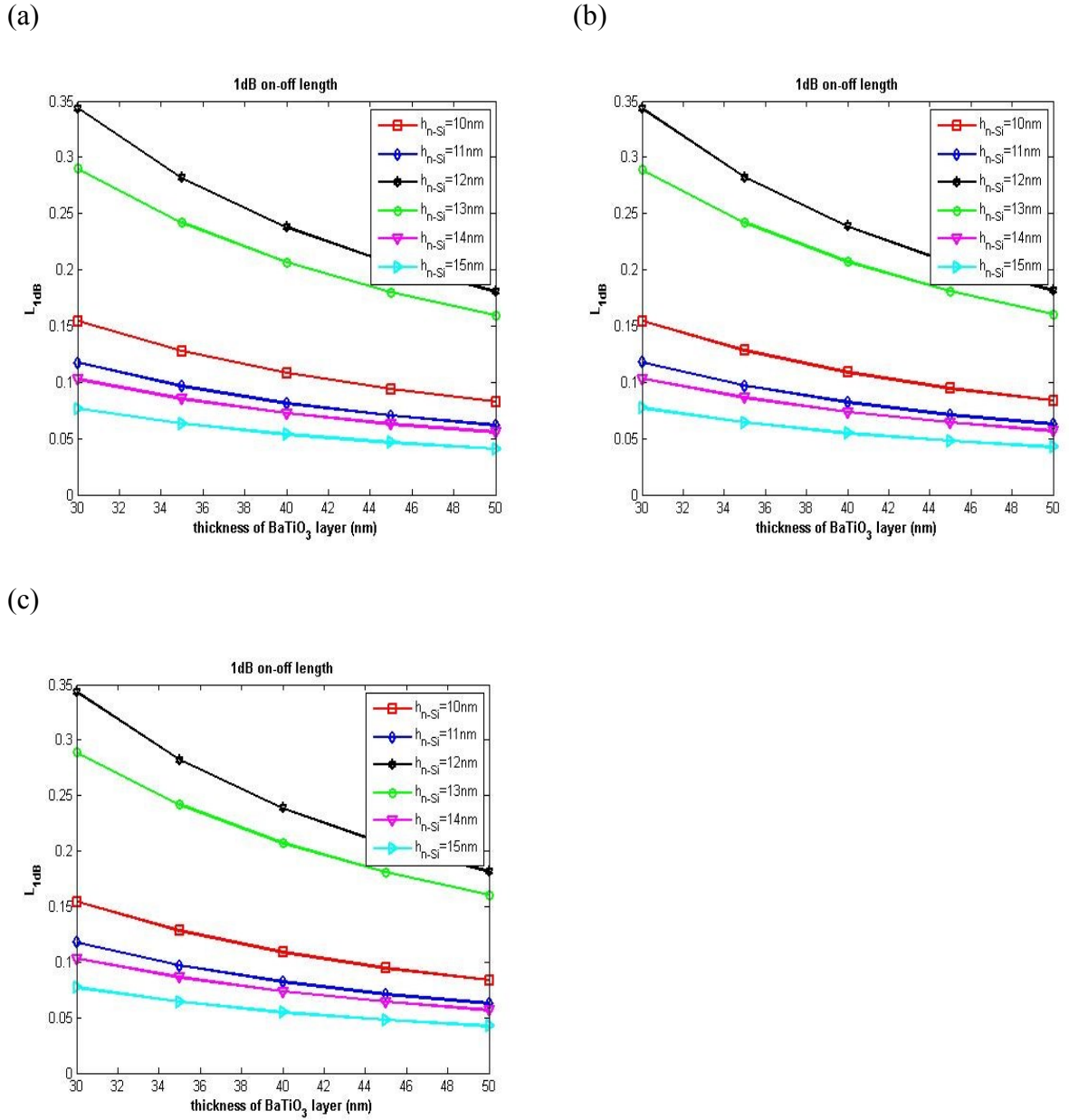

Figure 8. 1dB on-off length in (a) $h_{Au}$=30 nm, (b) $h_{Au}$=40 nm, (c) $h_{Au}$=50 nm.

The figure of merit in terms of the various layer thicknesses is depicted in Figure 9. It is obvious that with raising the $BaTiO_3$ thickness, the figure of merit decreases. Also, the maximum value of the figure of merit is 10.74 which is obtained at an Air($\infty$)/Au(50nm)/$BaTiO_3$(30nm)/n-Si(13nm)/Au(50nm)/$Al_2O_3$($\infty$) multilayer configuration. It should be noted that this value is higher than the obtained FoM value in the reference [36]. Furthermore, by increasing the thickness of the gold layer, the figure of merit raises.

(a)

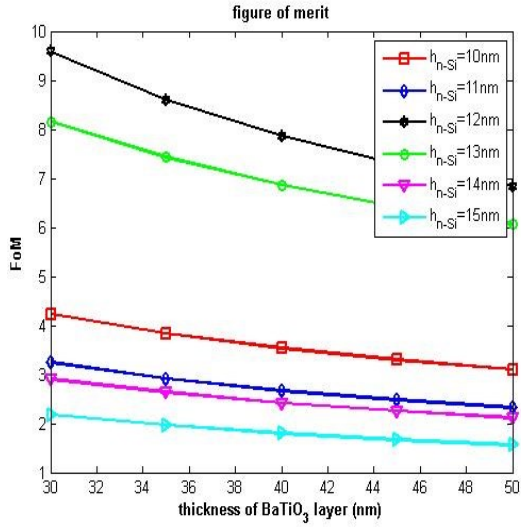

(b)

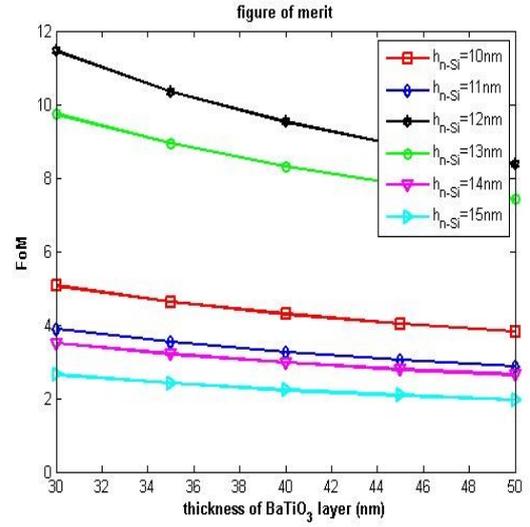

(c)

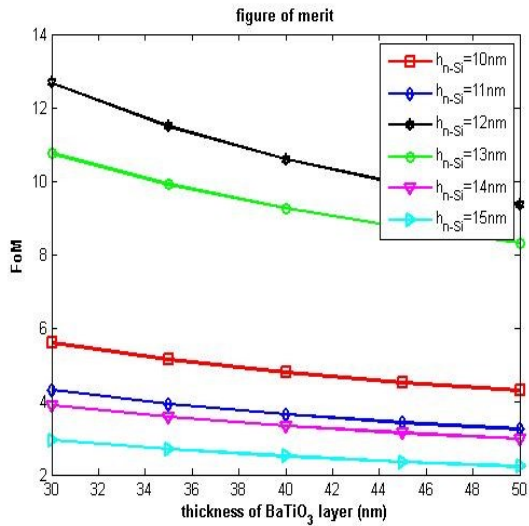

Figure 9. figure of merit in (a) $h_{Au}$=30 nm, (b) $h_{Au}$=40 nm, (c) $h_{Au}$=50 nm.

The distribution of magnetic field, electric field and the time-averaged Poynting vector of the asymmetric plasmonic mode and the guided SPP dispersion relation at the maximum value of the figure of merit are shown in Figure 10 and 11, respectively. time-averaged Poynting vector is enhanced by the applied voltage. it causes an intense alteration in absorption coefficient and a big figure of merit.

(a) (b)

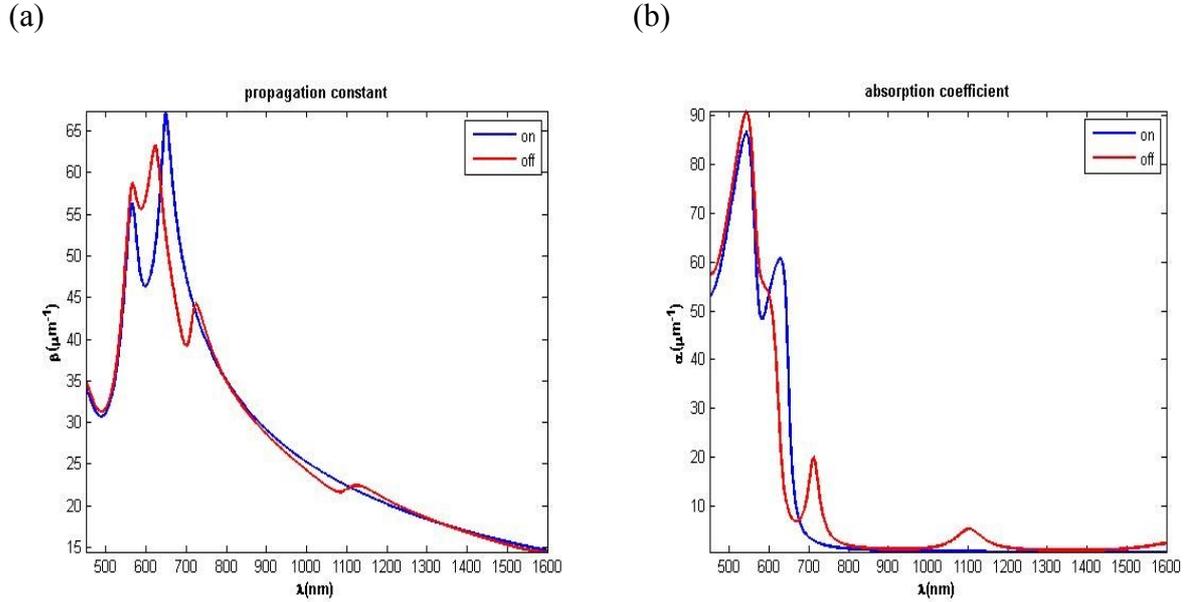

Figure 10. dispersion relation of the plasmonic mode in the Air($\infty$)/Au(50nm)/BaTiO$_3$(30nm)/n-Si(13nm)/Au(50nm)/Al$_2$O$_3$($\infty$) structure. (a) Propagation constant in the on and off states in different wavelengths, (b) absorption coefficient in the on and off states in different wavelengths.

(a) (b)

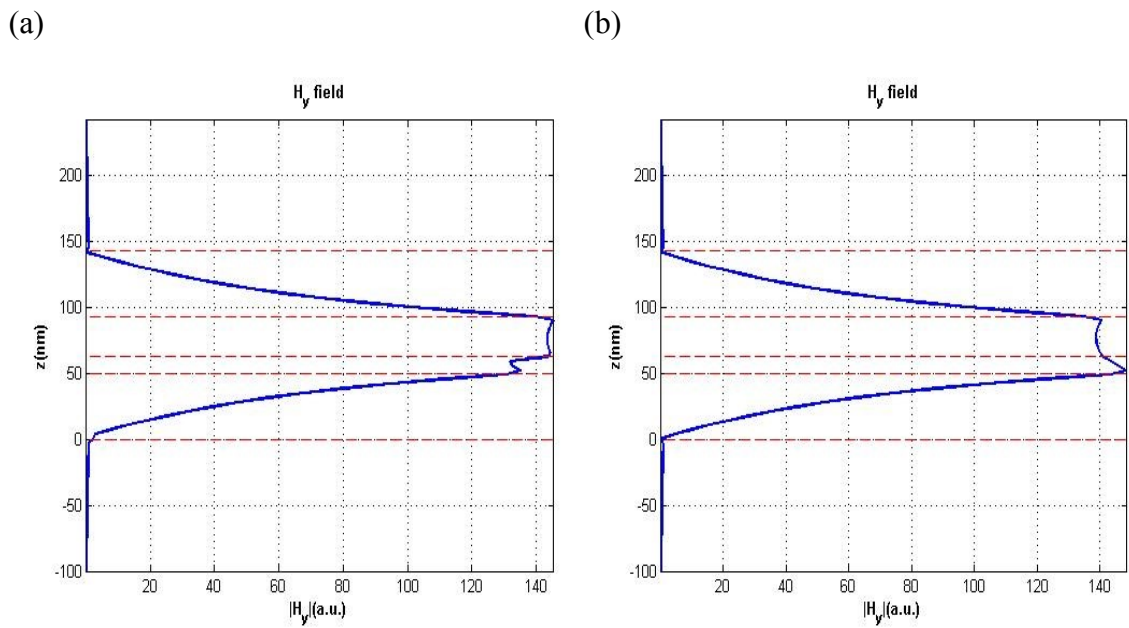

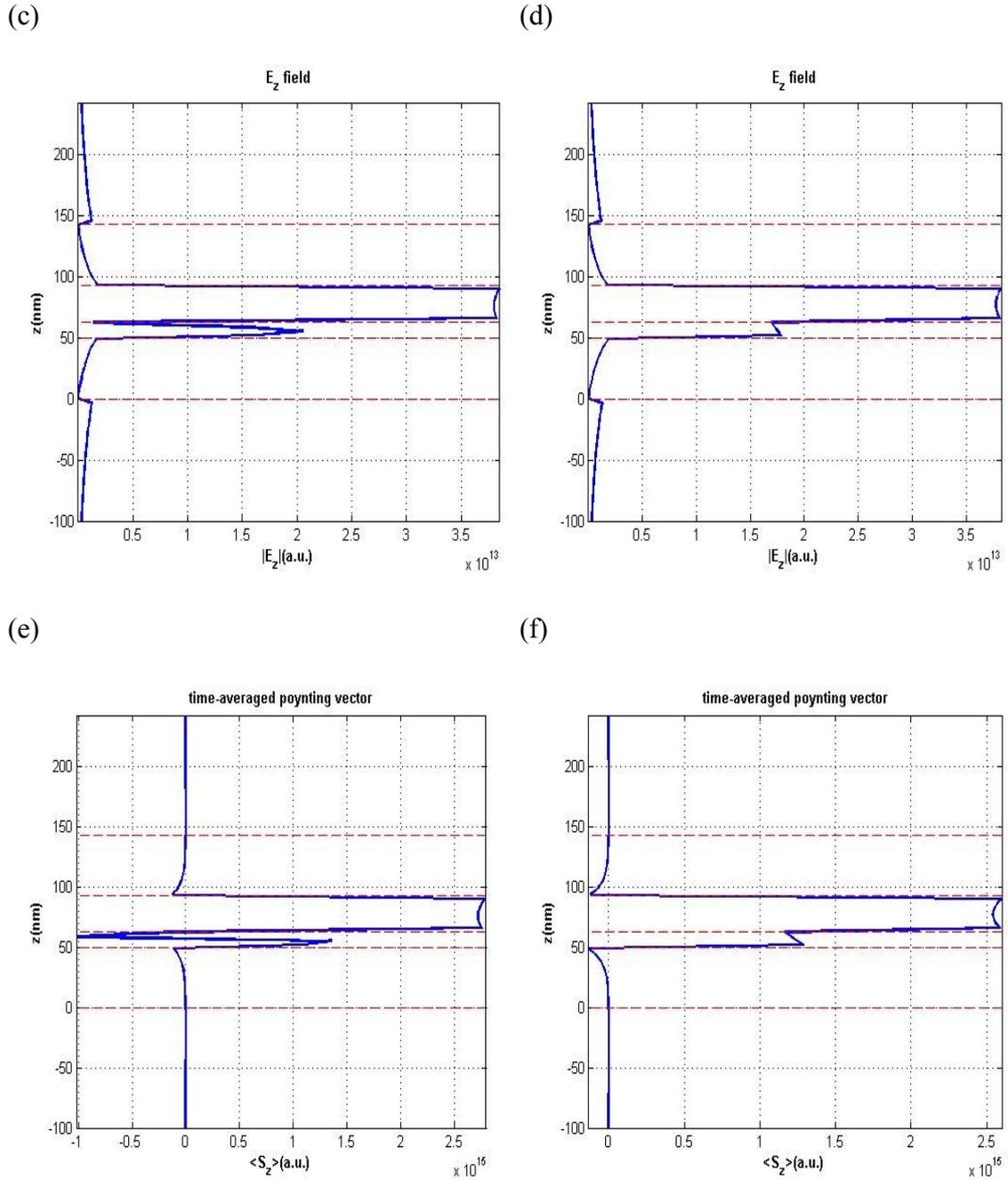

Figure 11. distribution of magnetic field in (a) the off and (b) on states and distribution of electric field in (c) the off and (d) on state and time-averaged Poynting vector in (e) the off and (f) on state of the Air(∞)/Au(50nm)/BaTiO3(30nm)/n-Si(13nm)/Au(50nm)/Al2O3(∞) structure.

## 4. Conclusion

In this study, a hybrid plasmonic modulator is introduced and its operational parameters are numerically examined. It is noteworthy that for more realistic simulation, the Pockels effect at the $BaTiO_3$ and the free carrier dispersion effect at the n-type silicon are considered simultaneously. Here, the related parameters of the phase and the absorption modulation are investigated as a function of different thicknesses. At the phase modulation viewpoint, it is shown that the minimum π shift length is equal to 6.91μm which happened at an Air(∞)/Au(50nm)/$BaTiO_3$(30nm)/n-Si(12nm)/Au(50nm)/$Al_2O_3$(∞) multilayer configuration. Also, the transmittance of this modulator is -15.36 and -60.16 in the on and off states, respectively. From the absorption modulation standpoint, the maximum value of the figure of merit, which occurs at an Air(∞)/Au(50nm)/$BaTiO_3$(30nm)/n-Si(13nm)/Au(50nm)/$Al_2O_3$(∞) multilayer structure, is equal to 10.74. By applying voltage, the energy flow increases in the n-Si layer and the barium titanate layer and it causes that the plasmonic mode is more sensitive for changing the refractive indexes of these two layers. According to our results, the discussed modulator is promising for exploiting as a phase and an intensity modulator, so it can be used in the plasmonic integrated circuits and also, due to compatibility with CMOS technology, it can be integrated with microelectronic systems.

**Acknowledgments**

We thank Abolfazl Safaei for assistance with editing this manuscript, and for his comments that greatly improved this paper and we thank Mehdi Shahsavari for his help in the editing of this manuscript.